\documentclass{svproc}
\usepackage{url}

\usepackage{graphicx}
\usepackage{setspace}
\begin{document}
\mainmatter              
\title{Towards new challenges of modern Pentest\thanks{A. F. Zorzo is supported by CNPq (grant 315192/2018-6). This study was financed in part by CAPES/Brazil - Finance Code 001. Also, we acknowledge the institutions that supported this study, namely CAPES, CNPq, and IFRS. Presented at World Conference on Smart Trends in Systems, Security, and Sustainability (WorldS4 2023)}}
\titlerunning{Towards new challenges of modern Pentest}  
%
\author{Daniel Dalalana Bertoglio\inst{1} \and Arthur Gil\inst{1} \and Juan Acosta\inst{1} \and Júlia Godoy\inst{1} \and Roben Castagna Lunardi\inst{2} \and Avelino Francisco Zorzo\inst{1} }
\authorrunning{Bertoglio et al.} 
%
%
\institute{Pontifical Catholic University of Rio Grande do Sul (PUCRS), Porto Alegre - RS, Brazil \and
Federal Institute of Rio Grande do Sul, Porto Alegre - RS, Brazil\\
\email{dalalana@gmail.com} \\
\email{\{a.gil, j.acosta, julia.godoy\}@edu.pucrs.br} \\
\email{roben.lunardi@restinga.ifrs.edu.br} 
\email{avelino.zorzo@pucrs.br}} 

\maketitle              

\begin{abstract}
With the increasing number of internet-based resources and applications, the amount of attacks faced by companies has increased significantly in the past years. Likewise, the techniques to test security and emulate attacks need to be constantly improved and, as a consequence, helping mitigate attacks. Among these techniques, penetration test (Pentest) provides methods to assess the security posture of assets, using different tools and methodologies applied in specific scenarios. Therefore, this study aims to present current methodologies, tools, and potential challenges applied to Pentest from an updated systematic literature review. As a result, this work provides a new perspective on the scenarios where penetration tests are performed. Also, it presents new challenges such as automation of techniques, management of costs associated with offensive security, and the difficulty in hiring qualified professionals to perform Pentest.

\end{abstract}
\section{Introduction}
Different domains and applications have emerged in the last years (\textit{e.g.}  Internet of Things, Smart Grids, Autonomous Cars, Smart Transportation), at the same time that cyberattacks have increased in complexity and quantity~\cite{Yang:2017,8611595:WorldS4,Musleh:2020}. Additionally, especially due to the pandemic, many companies have migrated their information/assets to the digital context. This migration implies new and different concerns for these companies, highlighting the importance of information security in terms of processes, technologies and new measures to face cyberattacks. Companies, organizations, and entities that suffer from these attacks need to find a way to mitigate the related risks. However, these companies are not able to deal with such risks in many situations (\textit{e.g.} new attacks, lack of testing in a system, software using deprecated code, etc.). As a consequence, risks that are not managed or mitigated can increase both the probability and impact of security attacks in the system, resulting in significant financial losses.


Assessing the security features of an asset is considered a essential task to manage its vulnerabilities. Security assessment is a continuous process to evaluate and recognize the risks that exist in a system~\cite{Mylrea:2018}. Usually, these security features evaluation is performed using security testing techniques. Therefore, applying correct techniques for this evaluation is important to reduce and minimize existing security risks in an organization \cite{ZHAO2010}.


Pentration test (Pentest) is one of most popular techniques to assess security of a application, system, infrastructure or a cyber-physical asset. Therefore, Pentest is a controlled process to explore a individual or multiple targets. As a result of the Pentest, is expected to identify and exploit vulnerabilities~\cite{McDermott:2001}. Also, it is designed to apply the same techniques employed in attacks performed by a malicious adversary. These penetration tests allow the development of a set of relevant tasks to identify and mitigate vulnerabilities before they can be explored by unauthorized/malicious users \cite{WHITAKER2005}. In general, Pentest process can be performed based on following steps: passive and active reconnaissance; attack surface mapping; exploiting found vulnerabilities; and post-exploration tasks~\cite{lam2004assessing}. Pentest can also be applied to understand if a company is properly performing the risk management and if the security process is comprehensive.


Due to the relevance of the security assessment, the objective of this work is to investigate and present a timeline of the evolution of the Pentest process. Hence, this work presents an update on the Systematic Mapping Study (SMS) that was performed previously by Bertoglio and Zorzo~\cite{DALALANA2017}. 
Additionally, this paper also identifies new research opportunities, frameworks, applications/scenarios, methodologies and tools in Pentest considering the 2017 to 2021. Regarding this update, the study also presents the main findings and presents and overview on the existing open issues in the Pentest. 

\section{Related Work} 
\label{sec:related}

The number of Mobile Cloud Computing (MCC) applications has significantly increased in the past years, and so did the complexity of its security testing. Al-Ahmad \textit{et al.} \cite{AlAhmad2019} presents a systematic literature review on penetration test in  MCC applications. They analyse the Pentest process, its advantages and issues in this scenario. Furthermore, they introduce a list of requirements to test this type of environment and discuss the implementation on testing stages. 

Yurtseven and Bagriyanik \cite{yurtseven2020review} also mentioned Cloud Computing, showing how to measure vulnerabilities and what the main Pentest practices are in cloud environments. Their research collects and classifies vulnerabilities and attack vectors studies by cloud delivery models in order to understand the cloud services and their vulnerabilities. Additionally, the authors present a literature review on common vulnerabilities and threats that cloud users encounter and also analyses Pentest strategies in this scenario.

Ankele \textit{et al.} \cite{ankele2019} define requirements and recommendations of Industrial Internet of Things (IIoT/IoT) systems. Therefore, they explore the IoT scenario - the most common devices and components - and conduct an overview on threat modelling and Pentest tools. Their work can help to automate, among others, the Pentest process.
When it comes to IoT, Cristoffer \textit{et al.} \cite{cristoffer2019} address the following problem: the lack of attention given to the security of IoT devices and, therefore, the existence of few specific tools for testing in this environment. With this in mind, the study proposes a systematic set of intrusion tests that specifically help in the security assessment in environments with IoT devices, reducing the need to use auxiliary or adapted tools.

Vats \textit{et al.} \cite{vats2020} conduct a comparative review of the literature of the area, focusing on Pentest tools. They discuss different strategies and types of Pentest, such as internal and external, blind and double blind and targeted Pentest strategies and white, gray and black box Pentest types. 

The scope of the research is the main difference between our study and the related ones. Unlike Al-Ahmad \textit{et al.} \cite{AlAhmad2019} and, Yurtseven and Bagriyanik \cite{yurtseven2020review}, which are specifically on Cloud Computing, or Ankele \textit{et al.} \cite{ankele2019} and Cristoffer \textit{et al.} \cite{cristoffer2019}, that focus on IoT, our study covers a multitude of scenarios. Moreover, Vats \textit{et al.} \cite{vats2020} present only tools and issues. In our study we consider scenarios, methodologies and challenges, presenting an overall analyses of the area.


\section{Pentest SMS Methodology}
\label{sec:sms}

We performed a Systematic Mapping Study (SMS)~\cite{KITCHENHAM2007} identify and to investigate Pentest research area. In our study, we focused on Pentest and its characteristics, methods and scenarios. It is important to note that this study is based and expands a previous work \cite{DALALANA2017}. 
The goal of this research is to provide an overview on Pentest tools, process and its general structure. We expect with this SMS to provide a comprehensive analysis that my help security professionals through the evolution of Pentest.
We defined the same Research Questions (RQ) presented in a previous work \cite{DALALANA2017}: \textbf{RQ1}: \textit{What are the main tools used in Pentest}; \textbf{RQ2}: \textit{What are the target scenarios in Pentest?}; \textbf{RQ3}: \textit{What are the models used in Pentest?}; and \textbf{RQ4}:\textit{ What are the main challenges in Pentest?}.





We selected ACM Digital Library, IEEE Xplore, Scopus and Springer Link as our databases. Considering our goal, we searched paper published from 2016 to 2021.
We defined our structured therms based on the research questions, resulting in the final search string: 

\begin{flushright}
\small{\texttt{(("penetration test" OR "penetration testing" OR pentest) AND (tool  OR tools OR  software OR suite) AND (model OR process OR framework OR methodology OR standard) AND (environment OR \\context) AND ("open research topics" OR challenges OR \\"open problems")).}}
\end{flushright}




Inclusion (IC) and Exclusion Criteria (EC) are used to perform a more adequate selection of papers. The IC criteria are: (\textit{i}) the study presents one or more tools for Pentest and (\textit{ii}) it introduces a Pentest process, model, methodology or framework; The EC criteria are: (\textit{i}) the study is not focused on Pentest; (\textit{ii}) the study presents a methodology for Pentest, but does not provide explanation about its use and application; and
(\textit{iii}) the study does not evaluate or present outcomes. Also, we used the following Quality Criteria (QA) to investigate the pertinence of papers: \textbf{QA1}: \textit{Does the work present a improvement to Pentest field?}; \textbf{QA2}: \textit{Is there an evaluation or discussion about he adoption of tools and/or models for Pentest?}; \textbf{QA3}: \textit{Does the study discuss the adopted tools/models?}. Finally, we applied the following score for each QA: No = 0; Partly = 0.5; Yes = 1. Therefore, the final score of each paper can range from 0 to 3, classified as follows: 2.5 to 3 (excellent), 2 (very good), 1.5 (good), 1 (regular) and 0 to 0.5 (limited).

The selection process is divided in four main steps: the first one refers to the search in data base through the search string. In the next one, the evaluation process is performed, according to the inclusion and exclusion criteria. The third step discusses quality assessment. Besides, a summary of each paper was prepared to assist in a future updating of this study. During these three steps, if the quality of a paper was not clear, there was a discussion among the three researchers. If no agreement was reached, another expert was consulted. The last step refers to handle the analysis result. In this step, all data is structured and a report of the study is prepared, finishing the systematic mapping process.

The management of a systematic mapping describe the process performed from the creation of the search string to the last steps of the analysis result. In this section, will be deepened the “Search on the data base” step and “Quality Assessment”. This initial search returned 1062 articles. Those articles were analysed through the inclusion and exclusion criteria. After that analysis, 1015 articles were eliminated, resulting in 47 articles that were thoroughly read.



The 47 selected papers were evaluated according the previously specified criteria.  After the Quality Assessment, resulted in 27 papers presented in the Table  \ref{tab: quality assessment}: the first column has the identification number (\textit{ID}); the second column contains the reference and published year; the \textit{Score} column shows the score obtained in this phase, and; the \textit{Description} column describes the classification.

\begin{table}[htb]
\centering
\caption{Quality Assessment Score}
\label{tab: quality assessment}
\scriptsize{
\begin{tabular}{cllcccccc}
\hline
      & \multicolumn{2}{c}{Papers}                      & \multicolumn{3}{c}{QA} & \multicolumn{2}{c}{Quality}  \\ \cline{2-8} 
ID    & \multicolumn{1}{l}{Study}                   & Year     & 1      & 2     & 3     & Score         & Description              \\ \hline
1     & \cite{applebaum2016}  Applebaum \textit{et al.}          & 2016     & Yes      & Yes     & Yes     & 3,0        & Excellent        \\
2     & \cite{morgner2017} Morgner \textit{et al.}               & 2017     & Yes      & Yes     & Yes     & 3,0        & Excellent        \\
3     & \cite{chung2016} Chung \textit{et al.}                   & 2016     & Yes      & Yes     & Yes     & 3,0        & Excellent        \\
4     & \cite{bhardwaj2020}  Bhardwaj \textit{et al.}            & 2020     & Yes      & Yes     & Yes     & 3,0        & Excellent        \\
5     & \cite{cristoffer2019} Cristoffer \textit{et al.}         & 2019     & Yes      & Yes     & Yes     & 3,0        & Excellent        \\
6     & \cite{pozdniakov2020} Pozdniakov \textit{et al.}         & 2020     & Yes      & Yes     & Yes     & 3,0        & Excellent        \\
7     & \cite{caselli2014} Caselli and Kargl            & 2016     & Yes      & Yes     & Yes     & 3,0        & Excellent        \\
8     & \cite{beckers2017}  Beckers \textit{et al.}              & 2017     & Yes      & Yes     & Yes     & 3,0        & Excellent        \\
9     & \cite{luh2020} Luh \textit{et al.}                       & 2020     & Yes      & Yes     & Yes     & 3,0        & Excellent        \\
10    & \cite{castiglione2020} Castiglione \textit{et al.}       & 2020     & Yes      & Yes     & Yes     & 3,0        & Excellent        \\
11    & \cite{ceccato2016} Ceccato and Scandariato      & 2016     & Partly      & Yes     & Yes     & 2,5        & Excellent        \\
12    & \cite{zheng2020} Zheng \textit{et al.}                   & 2020     & Yes      & Yes     & Partly     & 2,5        & Excellent        \\
13    & \cite{zhou2019} Zhou \textit{et al.}                     & 2019     & Yes      & Partly     & Yes     & 2,5        & Excellent        \\
14    & \cite{wang2016} Wang and Hong                   & 2016     & Partly      & Yes     & Yes     & 2,5        & Excellent        \\
15    & \cite{vondravcek2017} Vondr{\'a}\v{c}ek \textit{et al.}  & 2017     & Partly      & Yes     & Yes     & 2,5        & Excellent        \\
16    & \cite{salzillo2020} Salzillo \textit{et al.}             & 2020     & Yes      & Partly     & Partly     & 2,0        & Very Good        \\
17    & \cite{al2020}  Al-Ahmad \textit{et al.}                  & 2020     & Yes      & No     & Yes     & 2,0        & Very Good        \\
18    & \cite{guarda2016} Guarda \textit{et al.}                 & 2016     & Yes      & Partly     & Partly     & 2,0        & Very Good        \\
19    & \cite{falah2017}  Falah \textit{et al.}                  & 2017     & Partly      & Yes     & Partly     & 2,0        & Very Good        \\
20    & \cite{scully2018} Scully and Wang               & 2018     & Partly      & Partly     & Yes     & 2,0        & Very Good        \\
21    & \cite{patki2018} Patki \textit{et al.}                   & 2018     & No      & Partly     & Yes     & 1,5        & Good             \\
22    & \cite{ankele2019}  Ankele \textit{et al.}                & 2019     & Yes      & No     & Partly     & 1,5        & Good             \\
23    & \cite{ficco2017} Ficco \textit{et al.}                   & 2017     & Partly      & No     & Yes     & 1,5        & Good             \\
24    & \cite{al2018}  Al-Ahmad and Kahtan              & 2018     & Yes      & No     & Partly     & 1,5        & Good             \\
25    & \cite{relan2016} Relan                          & 2016     & Partly      & Partly     & Partly     & 1,5        & Good             \\
26    & \cite{oakley2019} Oakley                        & 2019     & Partly      & Partly     & Partly     & 1,5        & Good             \\
27    & \cite{antunes2017} Antunes and Vieira           & 2017     & Partly      & Partly     & Partly     & 1,5        & Good         
    \\ \hline
\end{tabular}
}
\end{table}

\section{Results and Discussion}
\label{sec:resultsanddiscussion}


This section answers each research question defined previously in Section \ref{sec:sms}.
To answer \textit{RQ1}, we analysed which tools were mentioned in the selected studies to verify the main tools used during Pentest, \textit{i.e.}, Metasploit, Nmap, Nessus, Burp Suite, SqlMap and IBM Rational AppScan.



Metasploit (IDs 6, 7, 10, 13, 16, 19, 20, 22 in Table \ref{tab: quality assessment}), Nmap (10, 14, 20, 22) and Nessus (6, 7, 13) remain as some of the main tools used in this area, since they were already reported in the 2017 paper~\cite{DALALANA2017}.Metasploit (open-source) can be easily customized and used with different operating systems. Also, in the past, Pentest required a lot of repetitive labor (\textit{e.g.} information gathering, gaining access and maintaining persistence), which now can be automated through Metasploit. Using this tool, a pentester can use ready-made or custom code to search for weak spots and vulnerabilities.

Nmap (a free and open-source tool) that can be used to identify network infrastructure, monitor hosts, verify services uptime, and manage services upgrades. It can perform fast scan on networks composed with hundreds of thousands of machines, but also works against single hosts. 

Finally, Nessus, a remote security scanning tool, scans a computer and alerts if it discovers any vulnerabilities that could be used to gain access to any computer connected to a network. As mentioned previously, Nessus is a scanning tool, therefore, it does not actively prevent attacks. Its purpose is to check computers for vulnerabilities. Nevertheless, most of the time it will be able to suggest the best way to mitigate the vulnerability - this is probably one of the reasons Nessus persists as one of the main tools in the Pentest area. Other reasons that keep it at the top are: it provides a scripting language to write specific tests to a system; also, it is possible to customize its source code; Nessus team updates the list of vulnerabilities on a daily basis, consequently the time window between the discovery of an exploit and Nessus being able to detect it is minimized.

We identified that 3 other tools have lately being widely used in Pentest: (\textit{i}) Burt Suite (IDs 11, 19, 22) is used for security testing and evaluation of web applications. Burp Suite can be used to perform automated scans for well-known web vulnerabilities. Additionally, it can offer manual testing to improve the Pentest; (\textit{ii}) SqlMap (IDs 19 and 22) is a penetration test tool that can be used to exploiting SQL databases with high level o automation. Also, it is used to recognize different Database Management System versions, which makes this tool a good alternative in the recognition and the gaining access phases; (\textit{iii}) IBM Rational AppScan (IDs 14 and 27), a web application security assessment suite that can be used to identify and fix common web application vulnerabilities. AppScan is a tool designed to test remote and local applications during the development process. The motivation is to reduce costs and fixing the vulnerabilites during the development phase.



To answer \textbf{RQ2}, this study shows that Pentest is applied in multiple scenarios. The recent development of Internet of Things (IoT)(IDs 5 and 16) increased the concern related to the security of such mechanisms, therefore Pentest applications in this area are a concern. In this scenario, Pentest acts in the identification of faults present in different systems in this environment, such as hardware, embedded systems, communication protocols and APIs. Reverse equipment engineering, memory dumps and cryptography analysis are part of a hardware-focused analysis of IoT, while port detection, impersonation overflow, detaining, and configuration of interfaces or backdoor, are focused on the firmware context of this scenario. Communication protocols such as NFC, ZigBee, Bluetooth, WiFi and Bluetooth Low Energy (BLE), can be explored through techniques such as analysis and multi-protocol capture of radio signals (``sniffing''), DoS, cryptography analysis, among several others. 

Regarding security tests on the Web (IDs 11, 12, 27) context, Pentest aims to exploit possible vulnerabilities in websites, applications, platforms, etc. Several security issues and vulnerabilities can be exploited in the Web context, \textit{e.g.} XSS, DDoS, SQL Injection, among others. In this sense, to perform the tests that can identify the possible vulnerabilities, different techniques and methodologies are adopted, \textit{e.g.} OWASP, OSSTMM, ISSAF, among others. Regarding Critical Infrastructure (CI) (IDs 7 and 23), a guide of stages is provided to identify and exploit vulnerabilities through methodologies and tools. However, in certain cases, Pentest can also be developed in a more humane than computational way. For example, it can be achieved using social engineering. 


Another recurrent scenario in the selected studies is Mobile Cloud Computing (IDs 17 and 24), which is a combination of Mobile Computing and Cloud Computing (ID 14). Cloud computing offers many benefits to mobile devices, such as data loss prevention, disaster recovery, and data security. However, Pentest in this type of scenario requires significant change. Particularly, each cloud type (IaaS, PaaS, or SaaS) requires different Pentest techniques and attack vectors.





\begin{figure*}[tb]
\centering
\includegraphics[width=1\textwidth]{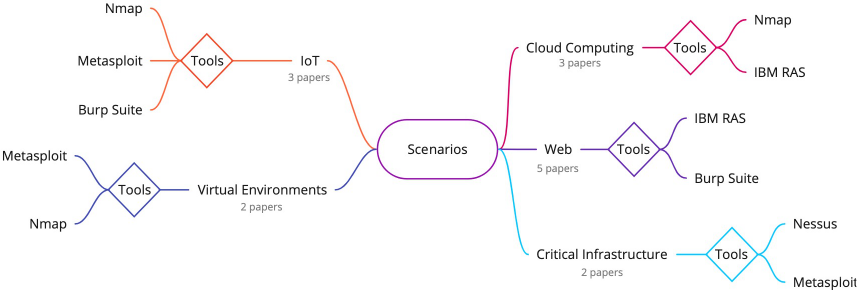}
\caption{Mind map on main Pentest scenarios and the respective tools}
\label{fig: mind map}
\end{figure*}

Figure \ref{fig: mind map} presents a relationship among the main Pentest scenarios and tools found in our SMS. Regarding the tools, we can observe that some of them appeared independently of the scenario. From this analysis, it was possible to highlight tools such as Metasploit, used in scenarios like IoT, Virtual Environment and Critical Infrastructure, and Nmap, with is used in IoT, Virtual Environment and Cloud Computing. Burp Suite, IBM RAS, and Nessus appeared in fewer scenarios compared to Metasploit and Nmap. 


\textbf{RQ3} covers the models and methodologies applied in Pentest. These methodologies establish the templates for conducting Pentest and act as a guide for professionals in the field. The main models obtained through the SMS are OWASP, OSSTMM, ISSAF, PTES and NIST SP 800-115. OSSTMM (IDs 7, 12, 16 and 21 in Table \ref{tab: quality assessment}) is a methodology that conducts tests on application security. Also,  Pentest in Cloud Computing can apply OSSTMM methodology. The ISSAF (IDs 7 and 12) is a framework used in security assessments, it provides a database of knowledge containing possible vulnerabilities of real-life scenarios. These two methodologies, generally, focus on guidelines and models for the evaluation of vulnerabilities and security risks. Considering that each Pentest procedure is unique to an environment, PTES (12, 16) provides a meticulous guideline for the process: basic standards, procedures and methods. Another methodology, OWASP (16, 21) centers on both web and mobile applications. Within OWASP there is the Top 10 project, Testing Guide, SAM and others that contribute to software security. Lastly, NIST SP 800-115 (7, 21) is also a document applied in system security. It gives an overview of specific security techniques, their benefits and limitations, and recommendations for their use. 



Regarding the new Pentest challenges (\textbf{RQ4}), both Pentest and the management of Red Team are expensive (monetary) processes~\cite{oakley2019}\cite{applebaum2016}. The general automation of the pentesting process is extremely complex and has yet to be developed. It is already known in advance that automation is a complicated subject to be addressed and even to be applied. For example, the authors Caselli and Kargl~\cite{caselli2014} approach this topic specifically in CI and Ankele et al.~\cite{ankele2019} discuss some automation concerns related to the IoT area. Most of the tools can fail to cover the human need in the main steps of Pentest, such as planning, preparation, conducting, and post-exploration. As presented, the amount of papers dealing with new scenarios that were not discussed in \cite{DALALANA2017} is significant. Thus, new challenges are present and new scenarios/technologies emerged.



\section{Comparative Analysis}

In our study, 49 Pentest tools were found among the selected papers. So, those that appeared at least 2 times were classified as the main ones. Table \ref{tab: comparison tools} shows the comparison between the main tools used in the Pentest area found in both studies. Through this analysis, it is possible to verify that 3 of the tools reported in the previous study \cite{DALALANA2017} are still part of Pentest's state of the art. The strengths of these tools were described in the previous section. 

\begin{figure*}[tb]
\centering
\includegraphics[width=1\textwidth]{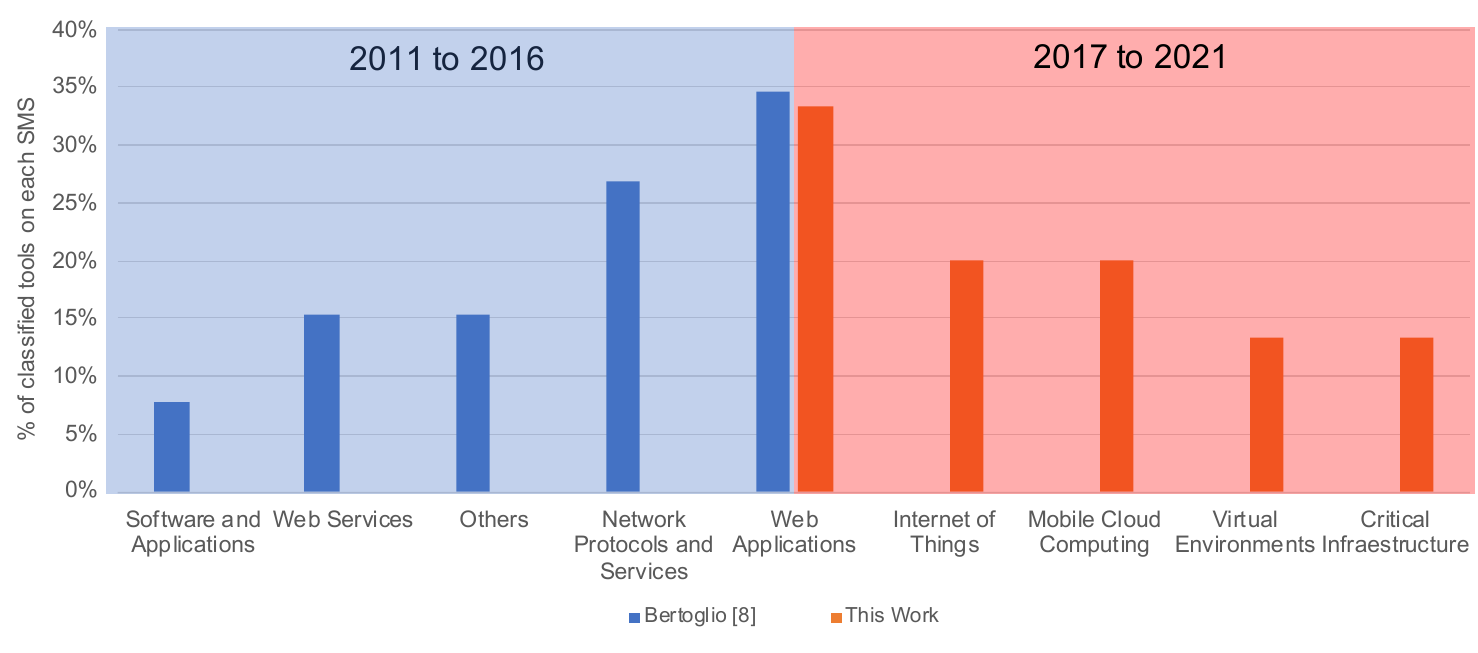}
\caption{Comparison of main scenarios in which tools are used}
\label{fig: comparison}
\end{figure*}

The study presented by Bertoglio and Zorzo \cite{DALALANA2017} shows that Web and Network Protocols and Services were the main scenarios in which Pentest was applied to. However, in our study, new scenarios emerged as presented in the previous section. 
This is a natural process, since Bertoglio and Zorzo  work\cite{DALALANA2017} was published 5 years before the current analysis; this time gap is enough for technology to evolve and require security validation in different areas.
Evidently, the results expressed on tables \ref{tab: comparison tools} and Figure \ref{fig: comparison} are linked, the tools that persisted over the years are those that managed to keep up with the evolution of scenarios. 

Regarding evolution on models, they have not changed significantly. Essentially, ``NIST Guidelines'' became ``NIST SP 800-115'' and ``OWASP Test Guide'' is simply referred now as ``OWASP''.

\begin{table}[tb]
\centering
\caption{Comparison: tools in 2011-2016 and 2017-2021}
\label{tab: comparison tools}
\scriptsize{
\begin{tabular}{cccl}
\hline
\multicolumn{4}{c}{Main tools}                                             \\ \hline
\multicolumn{2}{c|}{Bertoglio and Zorzo\cite{DALALANA2017} (2011-2016)}          & \multicolumn{2}{c}{\textbf{This work (2017-2021)}} \\ \hline
Acutenix   & \multicolumn{1}{c|}{NeXplore}    & \multicolumn{2}{c}{\textbf{Metasploit}}         \\
Metasploit & \multicolumn{1}{c|}{Nikto}       & \multicolumn{2}{c}{\textbf{Nmap}}               \\
Nessus     & \multicolumn{1}{c|}{Nmap}        & \multicolumn{2}{c}{\textbf{Nessus}}             \\
Paros      & \multicolumn{1}{c|}{QualysGuard} & \multicolumn{2}{c}{\textbf{Burp Suite}}         \\
WebInspect & \multicolumn{1}{c|}{WebScarab}   & \multicolumn{2}{c}{\textbf{SqlMap}}             \\
AppScan    & \multicolumn{1}{c|}{Wireshark}   & \multicolumn{2}{c}{\textbf{IBM Rational AppScan}}            \\ \hline
\end{tabular}
}
\end{table}

\section{Discussion \& Lessons Learned}
\label{sec:lessons}


Throughout this SMS, we analysed and compared several aspects not shown in our previous work\cite{DALALANA2017}. We can cite the differences between scenarios and the relation with different security tools adopted in several stages of Pentest. The main scenario in Pentest remains, mostly, the Web, as in the previous paper. However, there is currently a lot of discussion and studies in areas like mobile, IoT, and Cloud Computing, where, together, they appear in 6 of the 27 selected papers. While in the Bertoglio and Zorzo~\cite{DALALANA2017} there is an absence of studies regarding these areas, mainly IoT and Cloud Computing, which were already seen as a need for exploration by the authors.

Some new methodologies and tools were discussed. However, most remain the same when compared to the presented in Bertoglio and Zorzo~\cite{DALALANA2017}. This occurs due to existent methodologies and tools have already been accepted by the Pentest community. As these are already frequently used in several scenarios and stages of the Pentest, they are still used as a reference, not needing replacement.


One of the remaining concerns from the Bertoglio and Zorzo~\cite{DALALANA2017} is the need for test automation. The current automation only covers a small part of the entire Pentest process and it is precisely this issue that Caselli and Kargl \cite{caselli2014} approached by proposing a new methodology for security assessment in Critical Infrastructures (CIs). Their goal is to make tests more organized, reproducible and comparable, regarding tests in CIs. Regarding automation in IoT, Ankele\textit{ et al.}~\cite{ankele2019} developed a list of requirements and recommendations for these systems to include/adapt into their models. They also present a detailed analysis of IoT components and their desired parameters. These changes can further help in automating security analysis and Pentest in this field.


Furthermore, some of the selected papers describe new tools or frameworks that can be used during Pentest. Some of these tools present new strategies to help pentesters to find systems' vulnerabilities.
Some examples are: the framework for emulating Red Teams~\cite{applebaum2016}; the open-source Pentest framework, Z3sec, for testing the security of ZigBee 3.0 devices~\cite{morgner2017}; the PenQuest project, the RPG (Roleplaying Game) of APTs (Advanced Persistent Threats), which aims at a new teaching methodology through a video-game based on a multi-layered attacker/defender model that uses incentives through gamification~\cite{luh2020}.

\section{Conclusion}
\label{sec:conclusion}


We identified the Pentest application scenarios, tools, and methodologies used in different contexts from 2017 to 2021. Therefore, this SMS revisited and updated a previous work \cite{DALALANA2017}, using the same systematic mapping methodology. In the comparative analysis, we identified differences in scenarios, tools, challenges, and difficulties found in Pentest. There is high exploitation in Web, IoT, Cloud Computing, and CI environments, which are constantly targeted by attackers.

Some of the tools used in Pentest are well known and had little change over the years, in special to perform vulnerability scanning. In a comparison with the previous work, despite the evolution of technology, these tools continue to fulfill the needs and are effective in current Pentest. 
Likewise, models/methodologies applied to Pentest are still being adopted dominantly in the scope of Pentest. This is because such models are consolidated and adopted by pentesters.


Finally, automation persists as one of the most important challenges in Pentest. At the moment, there is no overarching framework that can test all types of systems. Another challenge is that systems and networks grew considerably in the past years, making it hard to execute manual Pentest. Also, there is a lack of Pentest qualified professionals. 
%
%
%
\bibliographystyle{splncs04}
\bibliography{mybibliography}
%




\end{document}